\newcommand*{\ATLASLATEXPATH}{}
\documentclass[cernpreprint, atlasdraft=false, texlive=2016, coverpage=false, UKenglish, subfigure]{\ATLASLATEXPATH atlasdoc}
 
\usepackage{\ATLASLATEXPATH atlaspackage}
\usepackage{\ATLASLATEXPATH atlasbiblatex}
 
\usepackage{\ATLASLATEXPATH atlasphysics}
\usepackage{\ATLASLATEXPATH atlasjetetmiss}

\graphicspath{{logos/}{figures/}}

\AtlasTitle{Measurement of the Lund jet plane using charged particles in 13~\TeV\ proton--proton collisions with the ATLAS detector}
 
\AtlasAbstract{
The prevalence of hadronic jets at the LHC requires that a deep understanding of jet formation and structure is achieved in order to reach the highest levels of experimental and theoretical precision. There have been many measurements of jet substructure at the LHC and previous colliders, but the targeted observables mix physical effects from various origins.  Based on a recent proposal to factorize physical effects, this Letter presents a double-differential cross-section measurement of the Lund jet plane using 139~fb$^{-1}$~of $\sqrt{s}=13$~\TeV\ proton--proton collision data collected with the ATLAS detector using jets with transverse momentum above 675~\GeV.  The measurement uses charged particles to achieve a fine angular resolution and is corrected for acceptance and detector effects.  Several parton shower Monte Carlo models are compared with the data. No single model is found to be in agreement with the measured data across the entire plane.}
 
\author{The ATLAS Collaboration}
 
\AtlasRefCode{STDM-2018-57}
 
\AtlasDate{\today}
 
\AtlasJournalRef{Phys. Rev. Lett. 124 (2020) 222002}
\AtlasDOI{10.1103/PhysRevLett.124.222002}
 
\PreprintIdNumber{CERN-EP-2020-030}
\hypersetup{pdftitle={ATLAS document},pdfauthor={The ATLAS Collaboration}}
 
\begin{document}
 
\maketitle
 
Jets are collimated sprays of particles resulting from high-energy quark and gluon production. The details of the process that underlies the fragmentation of quarks and gluons with quantum chromodynamic (QCD) charge into neutral hadrons is not fully understood. In the soft gluon (`eikonal') picture of jet formation, a quark or gluon radiates a haze of relatively low energy and statistically independent gluons~\cite{GATHERAL198390,FRENKEL1984231}.  As QCD is nearly scale-invariant, this emission pattern is approximately uniform in the two-dimensional space spanned by $\ln(1/z)$ and $\ln(1/\theta)$, where $z$ is the momentum fraction of the emitted gluon relative to the primary quark or gluon core and $\theta$ is the emission opening angle.  This space is called the Lund plane~\cite{Andersson:1988gp}.  The Lund plane probability density can be extended to higher orders in QCD and is the basis for many calculations of jet substructure observables~\cite{Dasgupta:2013ihk,Frye:2017yrw,Dasgupta:2018nvj,Salam:2016yht}.
 
The Lund plane is a powerful representation for providing insight into jet substructure; however, the plane is not observable because it is built from quarks and gluons. A recent proposal~\cite{Dreyer:2018nbf} describes a method to construct an observable analog of the Lund plane using jets, which captures the salient features of this representation. Jets are formed using clustering algorithms that sequentially combine pairs of proto-jets starting from the initial set of constituents~\cite{Salam:2009jx}. Following the proposal, a jet's constituents are reclustered using the Cambridge/Aachen (C/A) algorithm~\cite{Dokshitzer:1997in,Wobisch:1998wt}, which imposes an angle-ordered hierarchy on the clustering history. Then, the C/A history is followed in reverse (`declustered'), starting from the hardest proto-jet.  The Lund plane can be approximated by using the softer (harder) proto-jet to represent the emission (core) in the original theoretical depiction. For each proto-jet pair, at each step in the C/A declustering sequence, an entry is made in the approximate Lund plane (henceforth, the `primary Lund jet plane' or LJP) using the observables $\ln{(1/z)}$ and $\ln{(R/\Delta R)}$, with
\[
z = \frac{\pt^{\mathrm{emission}}}{\pt^{\mathrm{emission}}+\pt^{\mathrm{core}}}\hspace{4mm}\text{and}\hspace{4mm}\; \Delta R^2 = (y_\text{emission}-y_\text{core})^2+ (\phi_\text{emission}-\phi_\text{core})^2,
\]
where \pt is transverse momentum,\footnote{ATLAS uses a right-handed coordinate system with its origin at the nominal interaction point (IP) in the center of the detector and the $z$-axis along the beam pipe. The $x$-axis points from the IP to the center of the LHC ring, and the $y$-axis points upward. Cylindrical coordinates ($r$, $\phi$) are used in the transverse plane, $\phi$ being the azimuthal angle around the beam pipe. The pseudorapidity is defined in terms of the polar angle $\vartheta$ as $\eta = -\ln \tan(\vartheta/2)$.} $y$ is rapidity, $R$ is the jet radius parameter, and $\Delta R$ measures the angular separation. Using this approach, individual jets are represented as a set of points within the LJP. Ensembles of jets may be studied by measuring the double-differential cross section in this space. The substructure of emissions, which may themselves be composite objects, is not considered in this analysis. To leading-logarithm (LL) accuracy, the average density of emissions within the LJP is uniform~\cite{Dreyer:2018nbf}:
\begin{equation}\label{eqn:ll}
\frac{1}{N_{\mathrm{jets}}} \frac{d^{2}N_{\mathrm{emissions}}}{d\ln(1/z) d\ln(R/\DeltaR)} \propto \text{constant},
\end{equation}
where $N_{\mathrm{jets}}$ is the number of jets. This construction of the plane is selected to separate momentum and angular measurements, although other choices such as $\left(\ln(R/\Delta R),k_{t} = z\Delta R\right)$ are valid.
 
The Lund plane has played a central role in state-of-the-art QCD calculations of jet substructure~\cite{Frye:2016aiz,Frye:2016okc,Marzani:2017mva,Marzani:2017kqd,Kang:2018vgn,Kang:2018jwa} which have so far only been studied with the jet mass $m_\text{jet}$~\cite{STDM-2017-04,CMS-SMP-16-010} (which is itself a diagonal line in the LJP: $\ln{1/z} \sim \ln{m_\text{jet}^{2}/\pt^{2}} - 2\ln{R/\DeltaR}$) and groomed jet radius~\cite{Aad:2019vyi,Adam:2020kug}. The number of emissions within regions of the LJP is also calculable and provides optimal discrimination between quark and gluon jets~\cite{Frye:2017yrw}.
 
This Letter presents a double-differential cross-section measurement of the LJP, corrected for detector effects, using an integrated luminosity of 139~\ifb\ of $\sqrt{s} = 13~\TeV$ proton--proton ($pp$) collision data collected by the ATLAS detector. A unique feature of this measurement is that contributions from various QCD effects such as initial-state radiation, the underlying event and multi-parton interactions, hadronization, and perturbative emissions are well-localized in the LJP. This factorization is shown in Figure~\ref{fig:intro:schematic}, which qualitatively indicates the regions populated by soft \emph{vs}.\ hard, wide-angle \emph{vs}.\ collinear, and perturbative \emph{vs}.\ nonperturbative radiation. Since different regions are dominated by factorized processes, the LJP measurement can be useful for tuning nonperturbative models and for constraining the model parameters of advanced parton shower (PS) Monte Carlo (MC) programs~\cite{Fischer:2016vfv,Hoche:2015sya,Nagy:2015hwa,Bellm:2015jjp}.
 
The ATLAS detector~\cite{PERF-2007-01, ATLAS-TDR-2010-19, Abbott:2018ikt} is a general-purpose particle detector which provides nearly $4\pi$ coverage in solid angle. The inner tracking detector (ID) is inside a 2~T magnetic field and measures charged-particle trajectories up to $|\eta|=2.5$.  The innermost component of the ID is a pixelated silicon detector with fine granularity that is able to resolve ambiguities inside the dense hit environment of jet cores~\cite{PERF-2015-08}, surrounded by silicon strip and transition radiation detectors. Beyond the ID are electromagnetic and hadronic calorimeters, from which topologically connected clusters of cells~\cite{PERF-2014-07} are formed into jets using the \akt algorithm with radius parameter $R=0.4$~\cite{Cacciari:2008gp,fastjet}. The jet energy scale is calibrated so that, on average, the detector-level jet energy is the same as that of the corresponding particle-level jets~\cite{PERF-2016-04}. 
 
Events are selected using single-jet triggers~\cite{TRIG-2016-01,TRIG-2012-01}.  The leading and subleading jets are used for the measurement and are required to satisfy  $\pt^\text{leading} > 675~\GeV$ and $\pt^\text{leading} < 1.5\times \pt^\text{subleading}$.  This jet-$\pt$ balance simplifies the interpretation of the final state in terms of a \TwoToTwo scattering process.  Both jets must be within the ID acceptance ($|\eta|<2.1$). About 29.5~million jets satisfy these selection criteria.
 
Particle-level charged hadrons and their reconstructed tracks are used for this measurement because individual particle trajectories can be precisely identified with the ID. As the LJP observables are dimensionless and isospin is an approximate symmetry of the strong force, the difference between the LJP observables constructed using all interacting particles and charged particles is small~\cite{Aad:2019vyi}. Tracks are required to have $\pt > 500~\MeV$ and be associated with the primary vertex with the largest sum of track $\pt^2$ in the event~\cite{ATL-PHYS-PUB-2015-051}. Tracks within $\DeltaR=0.4$ of the cores of selected jets are used to construct the LJP observables by clustering them using the C/A algorithm and populating the plane by iterative declustering. The fiducial region of the measurement spans 19 bins in $\ln(1/z)$ between $\ln(1/0.5)$ and $8.4\times\ln(1/0.5)$, and 13 bins in $\ln(R/\DeltaR)$ between $0.0$ and $4.33$. The maximum $\Delta R$ is the jet radius and the minimum $\Delta R$ is comparable to the pixel pitch. The maximum $z$ is 0.5 and the minimum is $500 \text{ MeV} / p_\text{T}^\text{jet}$.
 
Samples of dijet events were simulated in order to perform the unfolding and compare with the corrected data.  The nominal sample was simulated using \PYTHIA~8.186~\cite{Sjostrand:2006za,Sjostrand:2007gs} with the NNPDF2.3 LO~\cite{Ball:2012cx} set of parton distribution functions (PDF), a \pt-ordered PS, Lund string hadronization~\cite{Andersson:1983ia,Sjostrand:1984ic}, and the A14 set of tuned parameters (tune)~\cite{ATL-PHYS-PUB-2014-021}. Additional samples were simulated by \PYTHIA~8.230~\cite{Sjostrand:2014zea} with the NNPDF2.3 LO PDF set and the A14 tune, using either the \PYTHIA LO matrix elements (MEs) or NLO MEs from \POWHEG~\cite{Nason:2004rx,Frixione:2007vw,Alioli:2010xa,Alioli:2010xd}; \SHERPA~2.1.1~\cite{Gleisberg:2008ta} with the CT10LO PDF set, a $p_\text{T}$-ordered PS~\cite{Schumann:2007mg}, an ME with up to three partons (merged with the CKKW prescription~\cite{Hoeche:2012yf}) and the \textsc{AHADIC} (A HADronization model In {C++}) cluster-based hadronization model~\cite{Webber:1983if,Winter:2003tt}; \SHERPA~2.2.5 with the CT14NNLO PDF set~\cite{Dulat:2015mca} including \TwoToTwo MEs and either the \textsc{AHADIC} hadronization model or the Lund string model; and \HERWIG~7.1.3~\cite{Bahr:2008pv,Bellm:2015jjp,Bellm:2017bvx} with the MMHT2014NLO PDF set~\cite{Harland-Lang:2014zoa} and either the default angle-ordered (Ang.~ord.) PS or a dipole PS and cluster hadronization~\cite{Webber:1983if}. Further details of these samples may be found in Ref.~\cite{ATL-PHYS-PUB-2019-017}. The \PYTHIA~8.186 and \SHERPA~2.1.1 events were passed through the ATLAS detector simulation~\cite{SOFT-2010-01} based on \GEANT~4~\cite{Agostinelli:2002hh}. The effect of multiple $pp$ interactions in the same and neighboring bunch crossings (pileup) was modeled by overlaying the hard-scatter event with minimum-bias $pp$ collisions generated by \PYTHIA~8 with the A3 tune~\cite{ATL-PHYS-PUB-2016-017} and the NNPDF2.3 LO PDF set. The distribution of pileup vertices was reweighted to match data, which have an average of $33.7$ simultaneous interactions per bunch-crossing.
 
Figures~\ref{fig:intro:ps}--\ref{fig:intro:me} illustrate the kinematic domains of various physical effects in the LJP using ratios at charged-particle level between pairs of MC simulations where one component of the simulation is varied. Varying the PS model in \HERWIG~7.1.3 (Figure~\ref{fig:intro:ps}) results in differences of up to 50\% in the perturbative hard and wide-angle emissions entering the lower-left region of the LJP. Changing the hadronization model in \SHERPA~2.1.1 (Figure~\ref{fig:intro:had}) causes variations up to 50\% in a different region of the plane, populated by softer and more collinear emissions at the boundary between perturbative and nonperturbative regions. Varying the ME from LO (\PYTHIA~8.230) to NLO (\POWPYTHIA~8.230) (Figure~\ref{fig:intro:me}) causes small changes of up to $10\%$ in the region populated by the hardest and widest-angle emissions.
 
\begin{figure}[hp]
\centering
\subfigure[Schematic representation of the LJP.]{\includegraphics[width=0.49\textwidth]{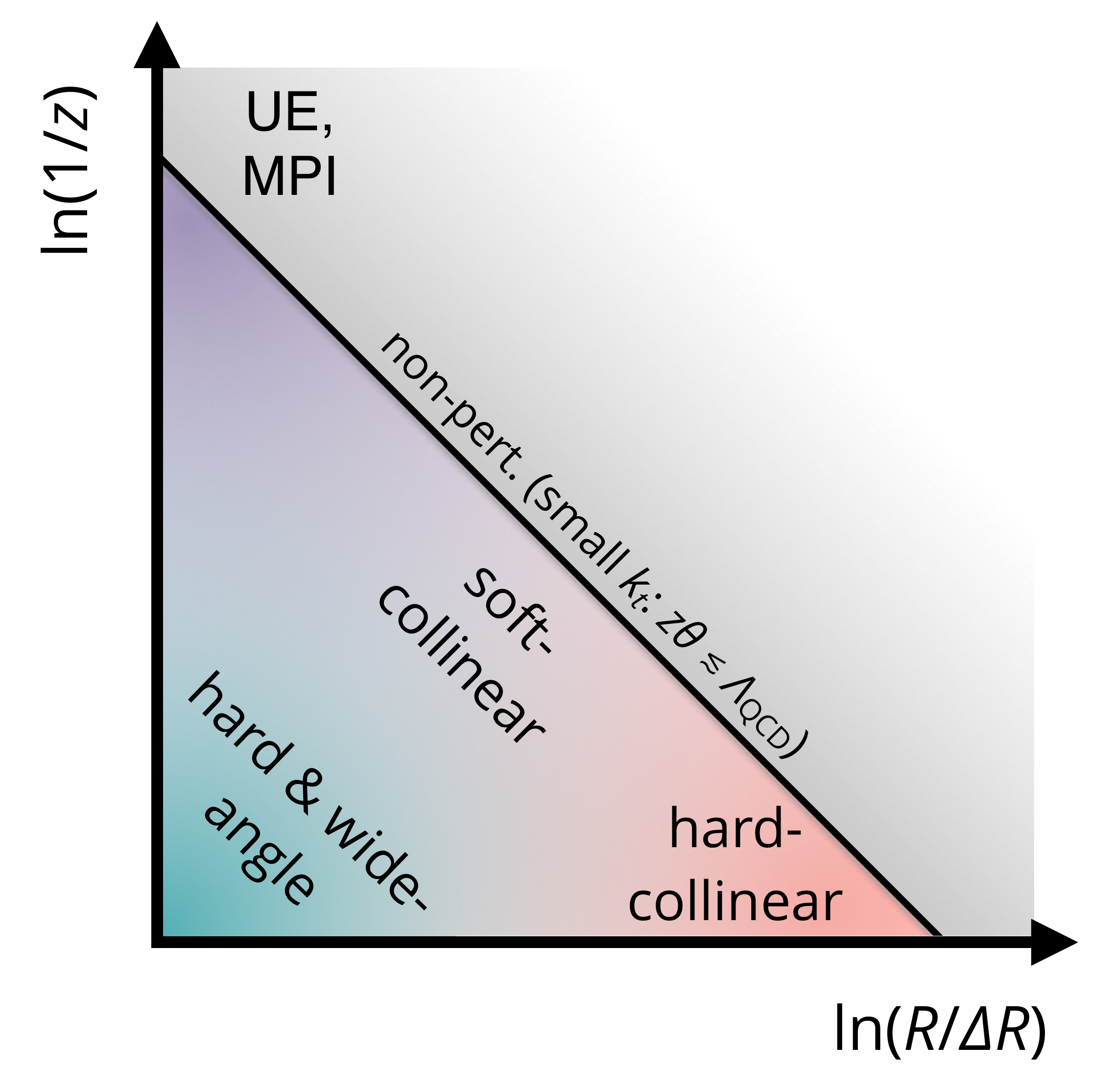}\label{fig:intro:schematic}}
\subfigure[Ratio of varied parton shower algorithms.]{\includegraphics[width=0.49\textwidth]{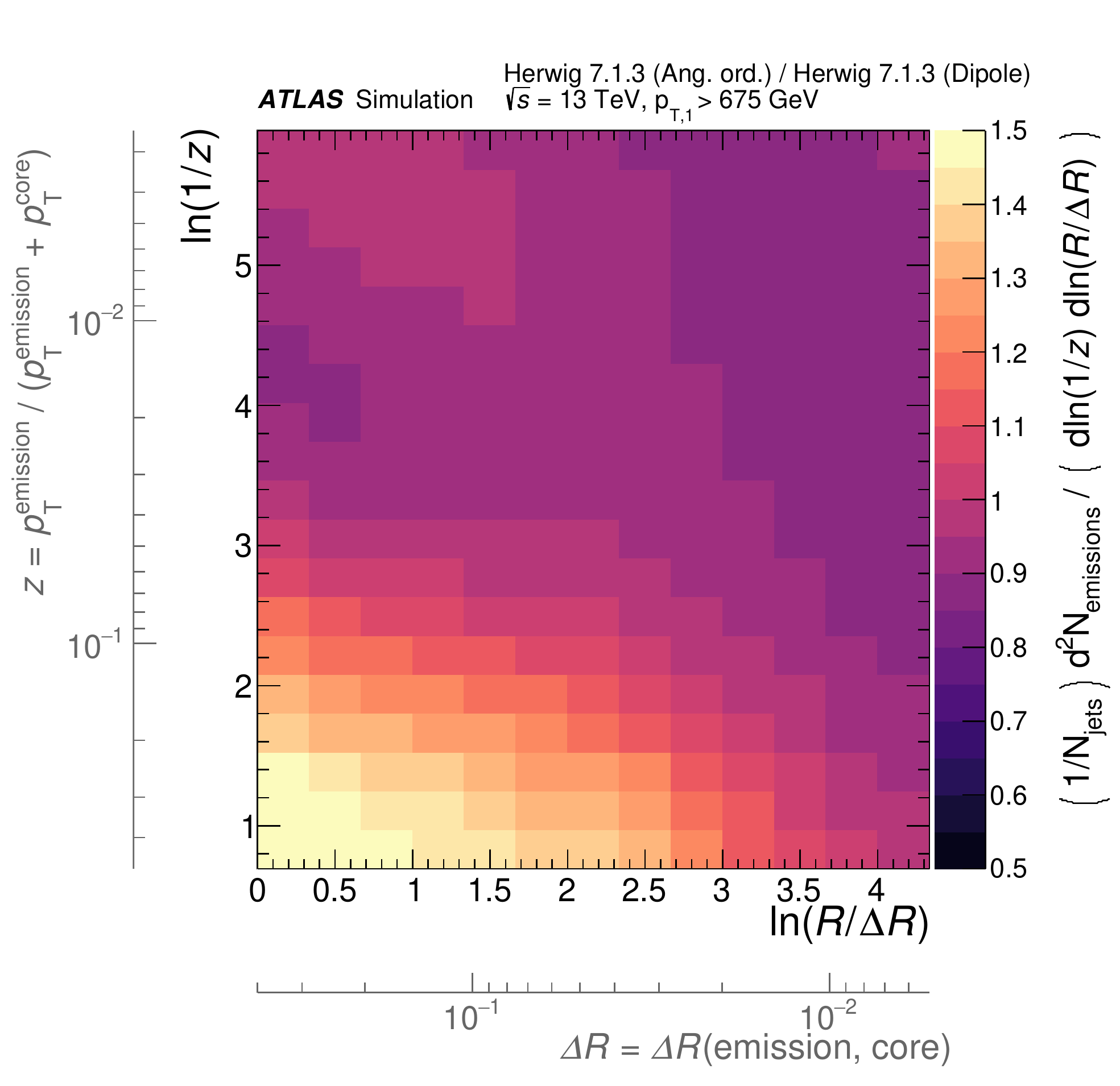}\label{fig:intro:ps}}	\\
\subfigure[Ratio of varied hadronization models.]{\includegraphics[width=0.49\textwidth]{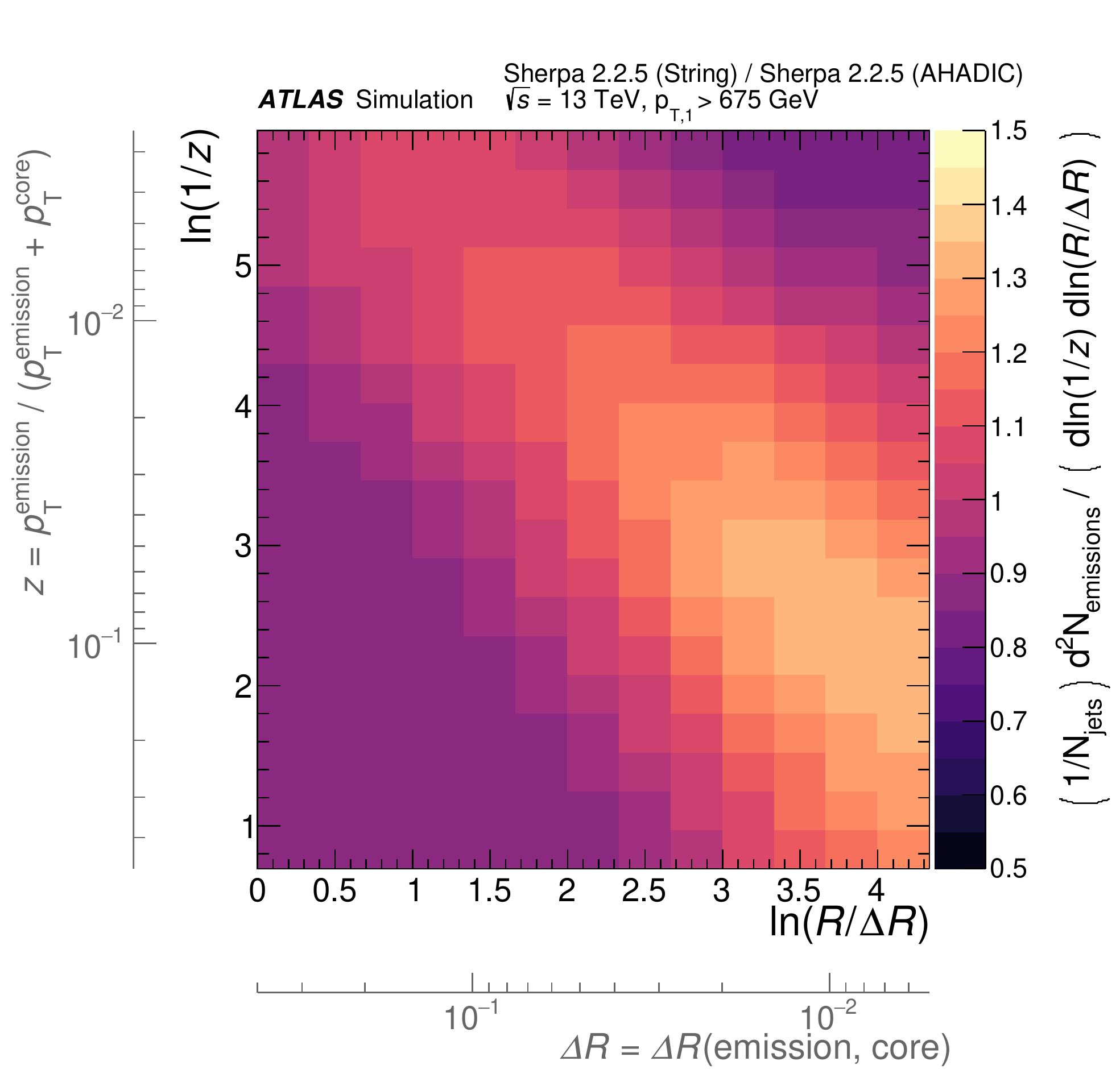}\label{fig:intro:had}}
\subfigure[Ratio of varied matrix elements.]{\includegraphics[width=0.49\textwidth]{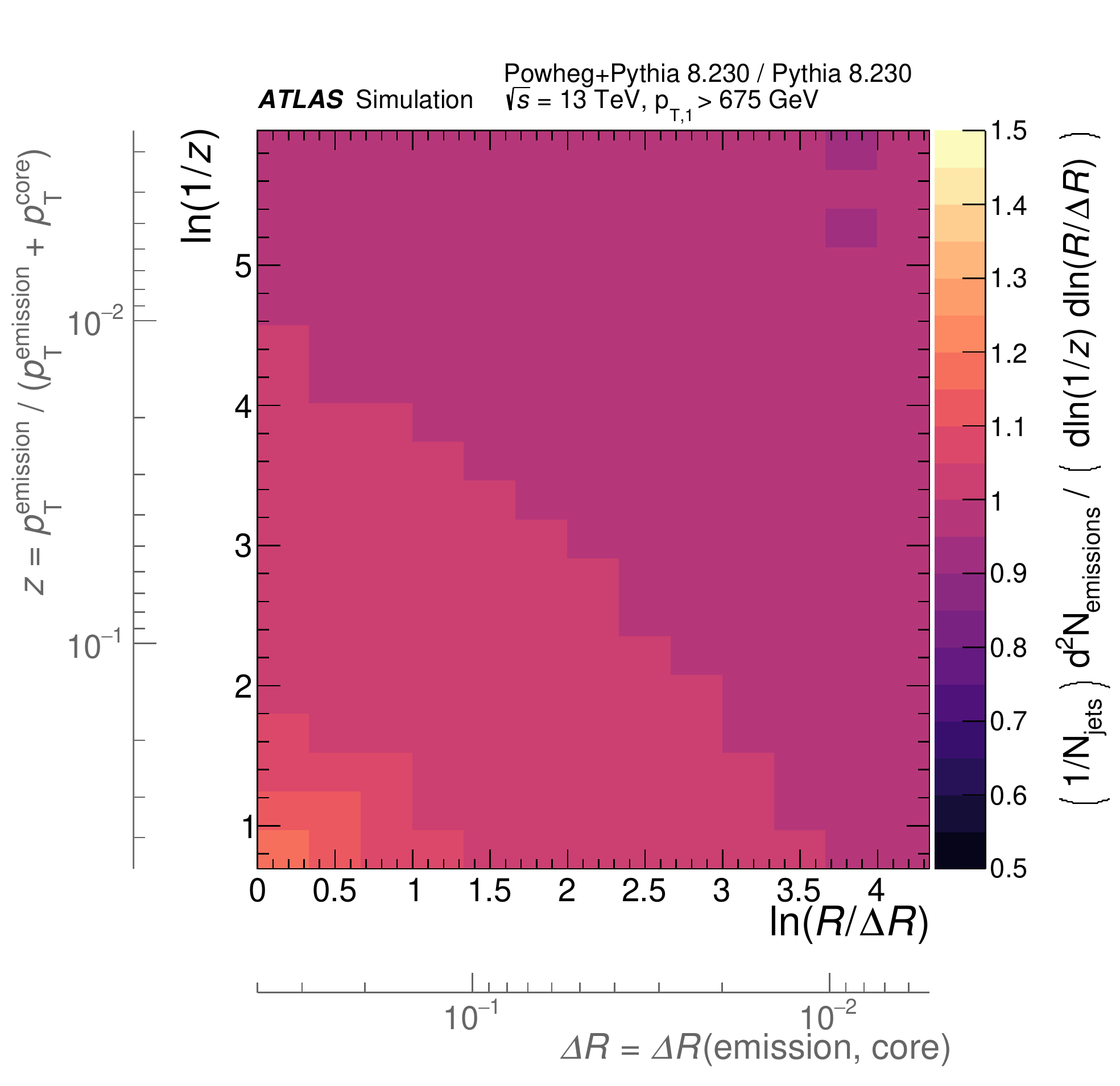}\label{fig:intro:me}}
\caption{
(a) Schematic representation of the LJP.  The line $z\theta \lesssim \Lambda_\text{QCD}$ roughly indicates the transition between regions where either perturbative ($z\theta>\Lambda_\text{QCD}$) or nonperturbative ($z\theta<\Lambda_\text{QCD}$) effects are expected to dominate. ``UE/MPI'' denotes the region where sources of nearly uniform radiation are relevant.  (b) The ratio of the Lund jet plane as simulated by the \HERWIG~7.1.3 MC generator with either an angle-ordered parton shower or a dipole parton shower. (c) The ratio of the Lund jet plane as simulated by the \SHERPA~2.2.5 MC generator with either the \textsc{AHADIC} cluster-based or Lund string-based hadronization algorithm. (d) The ratio of the LJP as simulated by either the \POWPYTHIA~8.230 or \PYTHIA~8.230 MC generators. The inner set of axes indicate the coordinates of the LJP itself, while the outer set indicate corresponding values of $z$ and $\DeltaR$.
}
\label{fig:intro}
\end{figure}
 
Selected data are unfolded to correct for detector bias, resolution, and acceptance effects by applying iterative Bayesian unfolding~\cite{DAgostini:1994fjx} with four iterations implemented in RooUnfold~\cite{Adye:2011gm}. The MC generator used to unfold the data is \PYTHIA 8.186. The number of iterations was chosen to minimize the total uncertainty. The unfolding procedure corrects the LJP constructed from detector-level objects to charged-particle level, where jets and charged particles are defined similarly to those at detector level: jets are reconstructed using the same anti-$k_t$ algorithm with detector-level stable ($c\tau > 10$~mm) non-pileup particles, excluding muons and neutrinos, as inputs.  The same kinematic requirements as for detector-level jets are imposed on these jets;  charged particles with $\pt > 500$~\MeV\ within $\Delta R = 0.4$ of the cores of particle-level jets are used to populate the charged-particle-level LJP.
 
Emissions at detector level and charged-particle level are uniquely matched in $\eta$$-$$\phi$ to construct the response matrix. The matching procedure follows the order of the C/A declustering, starting from the widest-angle detector-level emission and iterating towards the jet core. The closest charged-particle-level match with angular separation $\Delta R < 0.1$ takes precedence. Unmatched emissions from tracks not due to a single charged particle (detector level) and from nonreconstructed charged particles (charged-particle level) are accounted for with purity and efficiency corrections.  Corrections are applied before (purity) and after (efficiency) the regularized inversion of the response matrix. Both the purity and efficiency corrections are about 20\% for wide-angle, hard emissions (lower-left quadrant of the LJP), increasing to 80\% for the most collinear splittings and 50\% in the lowest-$z$ bins. For matched emissions, the $\ln(1/z)$ and $\ln(R/\Delta R)$ bin migrations between particle and detector levels are largely independent.  Furthermore, since the differential cross section varies slowly across the LJP, the purities and efficiencies are approximately the same across the entire LJP.  The $\ln(R/\Delta R)$ migrations in a given $\ln(1/z)$ bin are less than 60\% for the smallest opening angles and decrease to less than 40\% for the widest angles.  The $\ln(1/z)$ migrations decrease from about 50\% for the softest to about 20\% for the hardest emissions, with some degradation for the softest emissions at small opening angles. Migrations for both observables are nearly symmetric except for $\ln(R/\Delta R) > 3$, where harder-to-resolve small opening angles are measured with asymmetric resolution. In less than $10\%$ of these cases, particle-level and detector-level emissions are mismatched and therefore measured with the wrong $\ln(1/z)$. While the $\ln(R/\Delta R)$ migrations are nearly the same when $\ln(1/z)$ migrates by one bin, the $\ln(1/z)$ migrations increase by about 30\% when $\ln(R/\Delta R)$ migrates by one bin.
 
The unfolded distribution is normalized to the number of jets that pass the event selection, rendering the measurement insensitive to the total jet cross section. After normalization, the integral of the LJP is the average number of emissions within the fiducial region.
 
Experimental systematic uncertainties are evaluated by applying variations to each source, propagating them through the unfolding procedure, and taking the difference between the modified and nominal results. Theoretical uncertainties arise from jet fragmentation modeling. Different systematic uncertainties are treated as being independent. The size of various sources of uncertainty within selected regions of the LJP is displayed in Figure~\ref{fig:results:slices}.
 
Uncertainties in the jet energy are determined using a mixture of simulation-based and in situ techniques~\cite{PERF-2016-04}. These uncertainties cause the migration of jets into or out of the fiducial acceptance, and are typically above 3\% in total, reaching at most 7\%. Uncertainties related to the reconstruction of isolated tracks and tracks within dense environments are considered by modifying the measured \pt of individual tracks or removing them completely~\cite{PERF-2015-08,PERF-2015-07,PERF-2015-08}. These uncertainties are small, contributing less than $0.5\%$. Other experimental uncertainties related to the modeling of pileup and the stability of the measurement across data-taking periods are less than $1\%$ except for the most collinear splittings, where they reach $5\%$. A data-driven nonclosure uncertainty is determined by unfolding the detector-level distribution following a reweighting based on a comparison of the corresponding simulated detector-level distribution with the data~\cite{Malaescu:2009dm}.  This uncertainty is less than 1\% except for the most collinear splittings, where it approaches 5\%.  An uncertainty for the matching procedure between emissions at detector and charged-particle levels is determined by repeating the unfolding and iterating through the C/A declustering sequence in reverse (from collinear to wide-angle emissions), taking the change in the result as an uncertainty. This uncertainty is less than 1\% everywhere.
 
Theoretical uncertainties arise mainly from the accuracy of jet fragmentation modeling. Variations in jet fragmentation can impact the result through a combination of sources: efficiency/purity corrections, response matrix, and unfolding prior. These contributions are estimated by repeating the unfolding with \SHERPA~2.2.1. As the correlation between the uncertainty sources is unknown, an envelope of the 100\% and 0\% correlation hypotheses is taken as the total modeling uncertainty. This uncertainty ranges between 5\% and 20\% depending on the region (larger for soft-collinear splittings) and is the largest single source of uncertainty. Experimental uncertainties are found to be comparable to those arising from modeling in some regions of the LJP.
 
The total systematic uncertainty varies across the LJP; an uncertainty between 5\% and 20\% is achieved. The uncertainty is found to increase as $k_{t}=z\DeltaR$ decreases: the bin with the smallest $k_{t}$ is also measured least-precisely, and has a total uncertainty of about $20\%$.
 
The unfolded LJP is shown in Figure~\ref{fig:results:money}. A triangular region with $k_t \gtrsim \Lambda_\text{QCD}$ is populated nearly uniformly by perturbative emissions, agreeing with the LL expectation (Eq.~(\ref{eqn:ll})).  A large number of emissions are found at the transition to the nonperturbative regime, as $\alpha_\text{s}$ is enhanced for small values of $k_t$. Emissions beyond the transition fall within the nonperturbative region of the LJP ($k_t \lesssim \Lambda_\text{QCD}$), and are suppressed. The average number of emissions in the fiducial region is measured to be  $7.34~\pm~0.03~\text{(syst.)}~\pm~0.11~\text{(stat.)}$. The uncertainty is estimated by propagating uncertainties from the measurement in an uncorrelated and symmetrized manner. The corresponding average emissions for \PYTHIA~8.230 is 7.64 and 7.67 for \POWPYTHIA~8.230. The average value for \SHERPA~2.2.5 is 6.90 for AHADIC hadronization and 7.30 for Lund string hadronization. The average value for \HERWIG~7 is 7.41 for the dipole PS and 7.37 for the angle-ordered PS. While a similar bracketing of the data by \PYTHIA and \SHERPA with \textsc{AHADIC} hadronization was noted in Ref.~\cite{STDM-2015-12}, the particle multiplicity inside jets has not previously been decomposed into perturbative and nonperturbative components.
 
Figure~\ref{fig:results:slices} shows data from four selected horizontal and vertical slices through the LJP, along with a breakdown of the systematic uncertainties. The data are compared with predictions from several MC generators.  While no prediction describes the data accurately in all regions, the \HERWIG~7.1.3 angle-ordered prediction provides the best description across most of the plane. The differences between the PS algorithms implemented in \HERWIG~7.1.3 are notable at large values of $k_t = z\DeltaR$, where the two models disagree most significantly for hard emissions reconstructed at the widest angles (Figures~\ref{fig:results:slices:a} and \ref{fig:results:slices:b}). The \POWPYTHIA and \PYTHIA predictions only differ significantly for hard and wide-angle perturbative emissions, where ME corrections are relevant.  The hadronization algorithms implemented in \SHERPA~2.2.5 are most different at small values of $k_{t}$, particularly for soft-collinear splittings at the transition between perturbative and nonperturbative regions of the plane. The ability of the LJP to isolate physical effects is highlighted in Figure~\ref{fig:results:slices:b}, where as emissions change from wide-angled to more collinear, the distribution passes through a region sensitive to the choice of PS model, and then enters a region which is instead sensitive to the hadronization model. Figures~\ref{fig:results:slices:c} and \ref{fig:results:slices:d} show regions dominated by nonperturbative effects. The \PYTHIA samples describe the data in the collinear region of the jet core well, but all simulations fail to describe the softest, widest-angle emissions, which are characteristic of contributions from the underlying event. The \PYTHIA~8.186 and \SHERPA~2.2.1 predictions are not shown, but are consistent with the \PYTHIA~8.230 and \SHERPA~2.2.5 (Lund string hadronization) predictions, respectively. These observations indicate that the LJP may provide useful input to both perturbative and nonperturbative model development and tuning.
 
\begin{figure}[hp]
\centering
\includegraphics[width=0.95\textwidth]{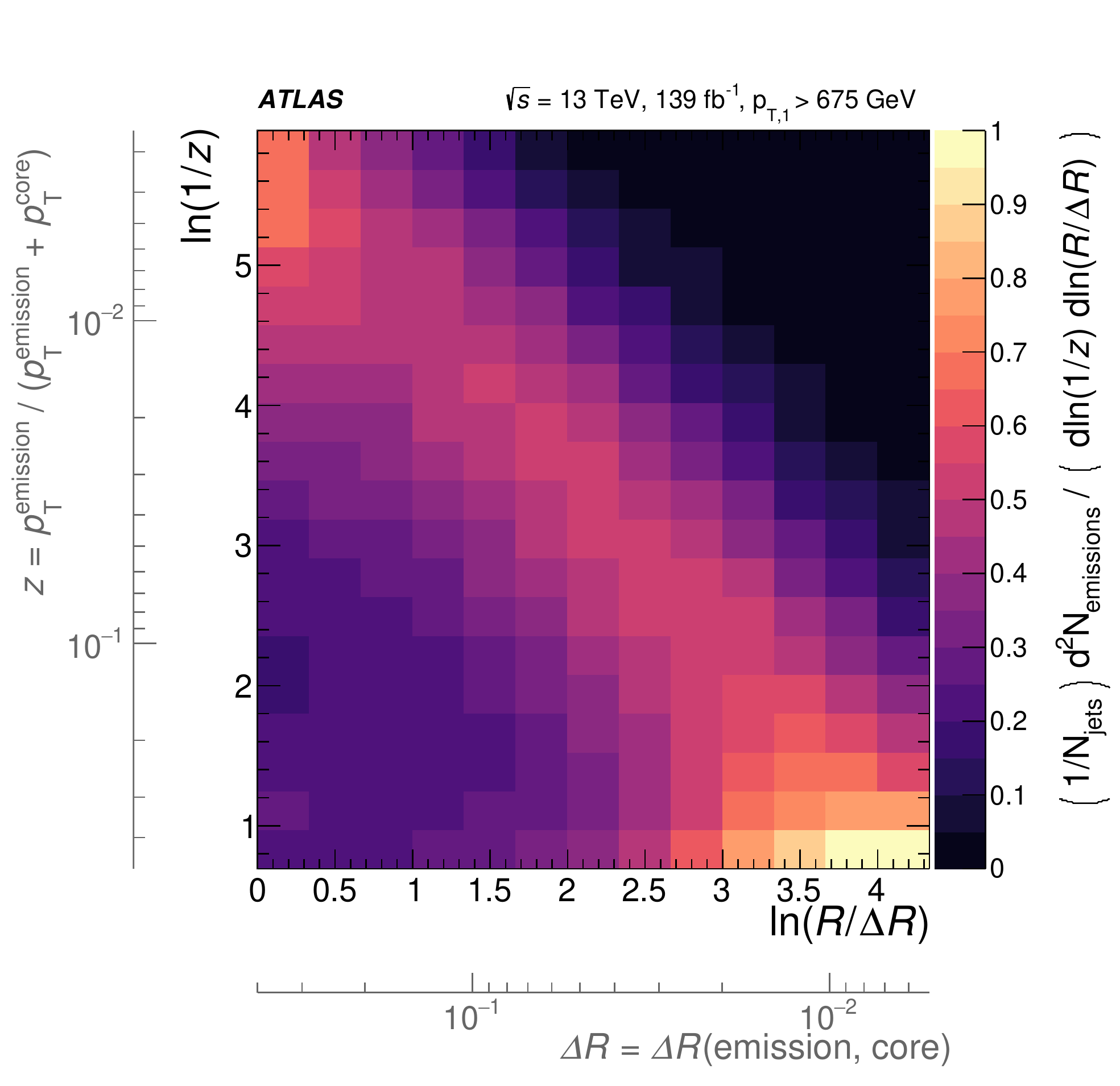}\\
\caption{
The LJP measured using jets in 13~\TeV~$pp$ collision data, corrected to particle level. The inner set of axes indicates the coordinates of the LJP itself, while the outer set indicates corresponding values of $z$ and $\DeltaR$.
}
\label{fig:results:money}
\end{figure}
 
\begin{figure}[ht]
\centering
\subfigure[]{\includegraphics[width=0.49\textwidth,trim=25 0 30 25,clip]{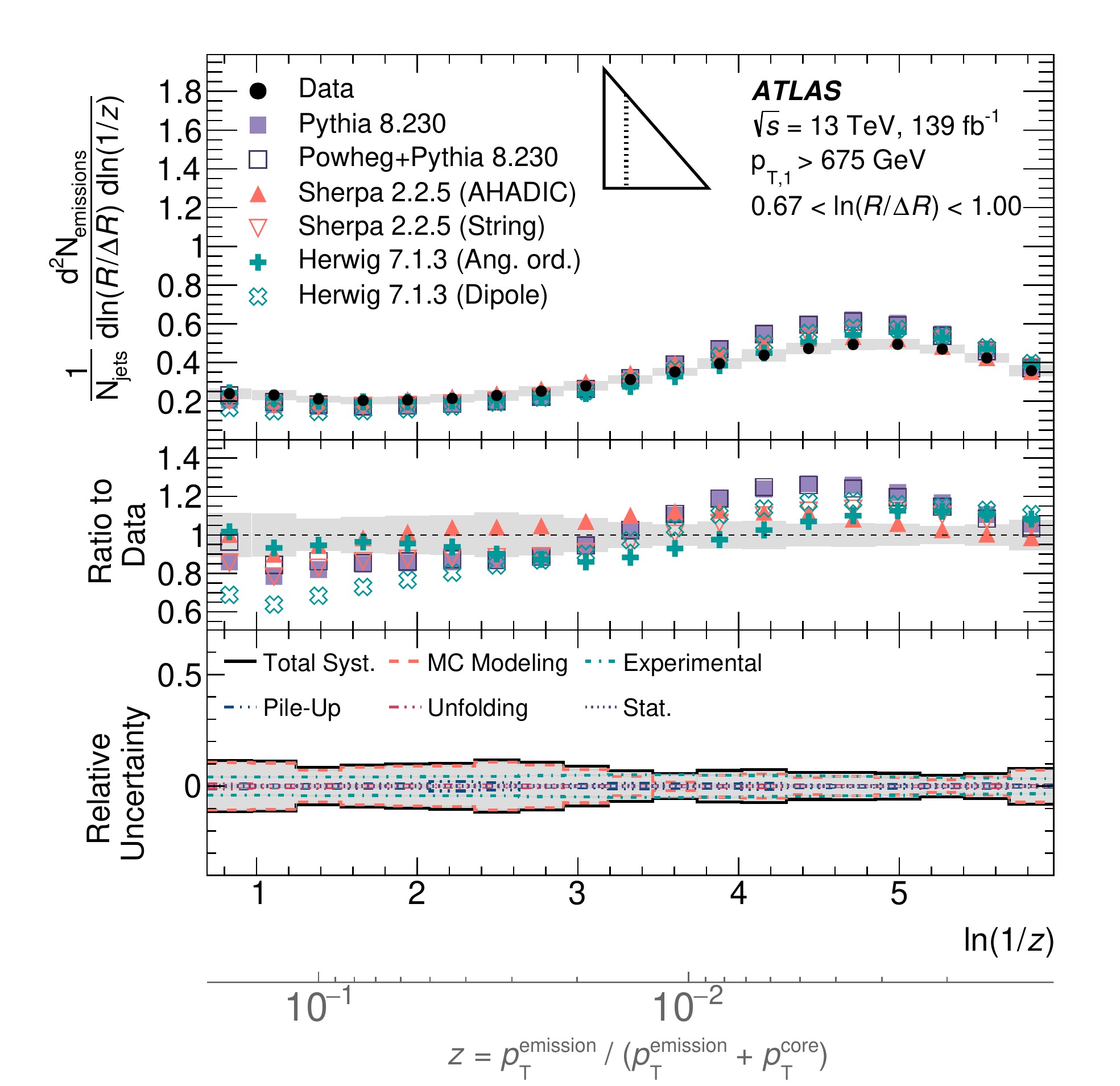}\label{fig:results:slices:a}}
\subfigure[]{\includegraphics[width=0.49\textwidth,trim=25 0 30 25,clip]{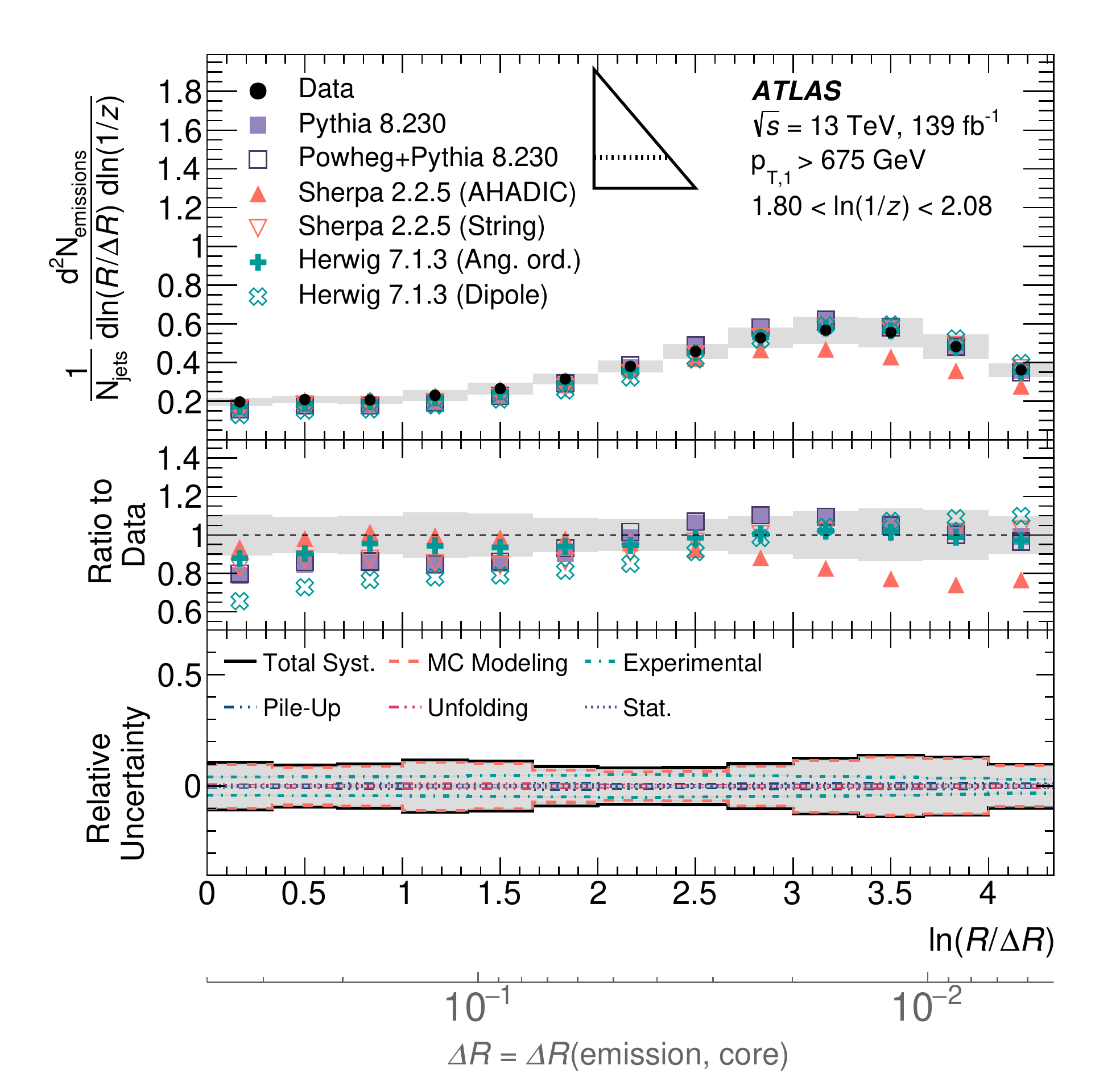}\label{fig:results:slices:b}}
\subfigure[]{\includegraphics[width=0.49\textwidth,trim=25 0 30 25,clip]{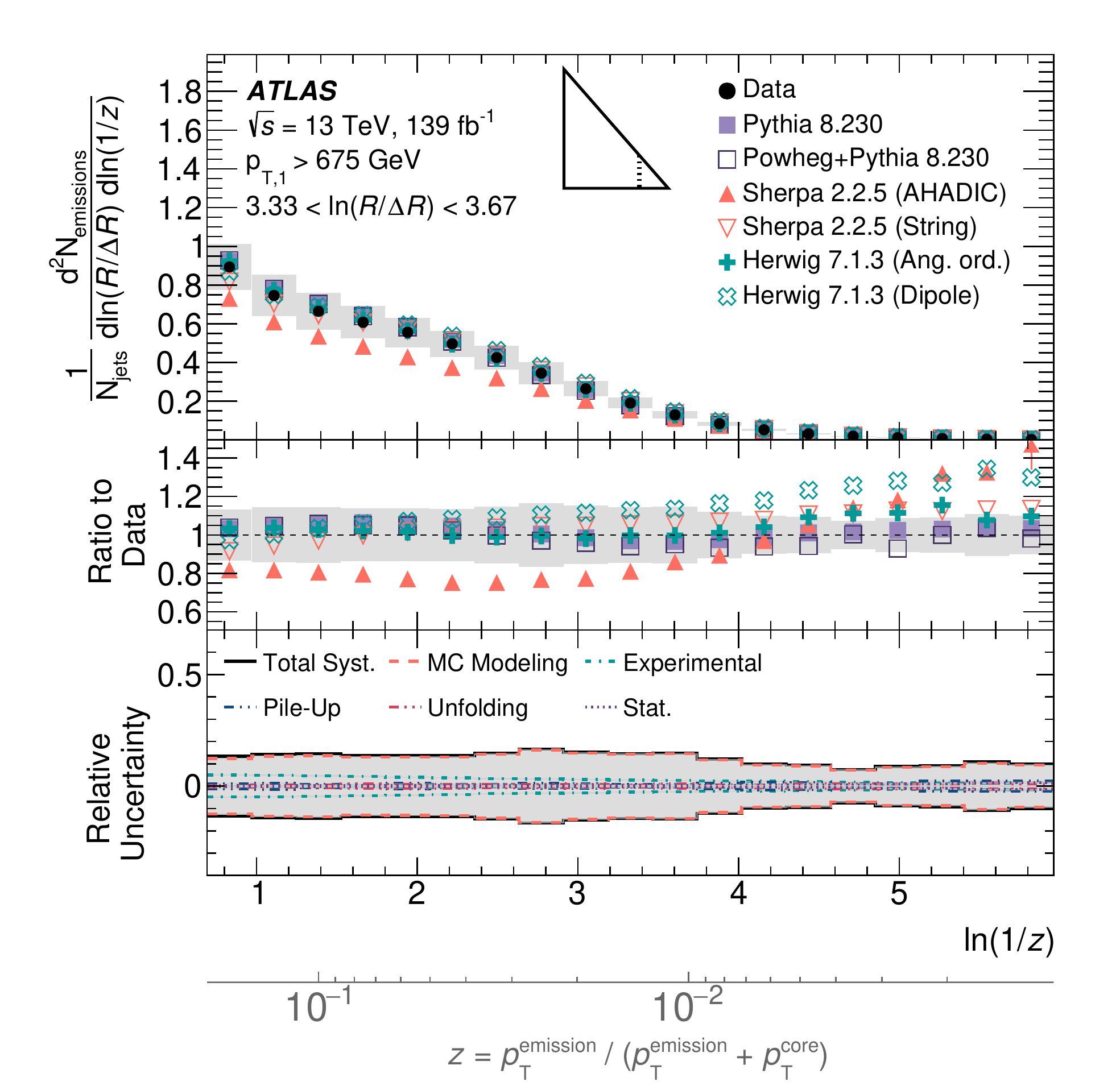}\label{fig:results:slices:c}}
\subfigure[]{\includegraphics[width=0.49\textwidth,trim=25 0 30 25,clip]{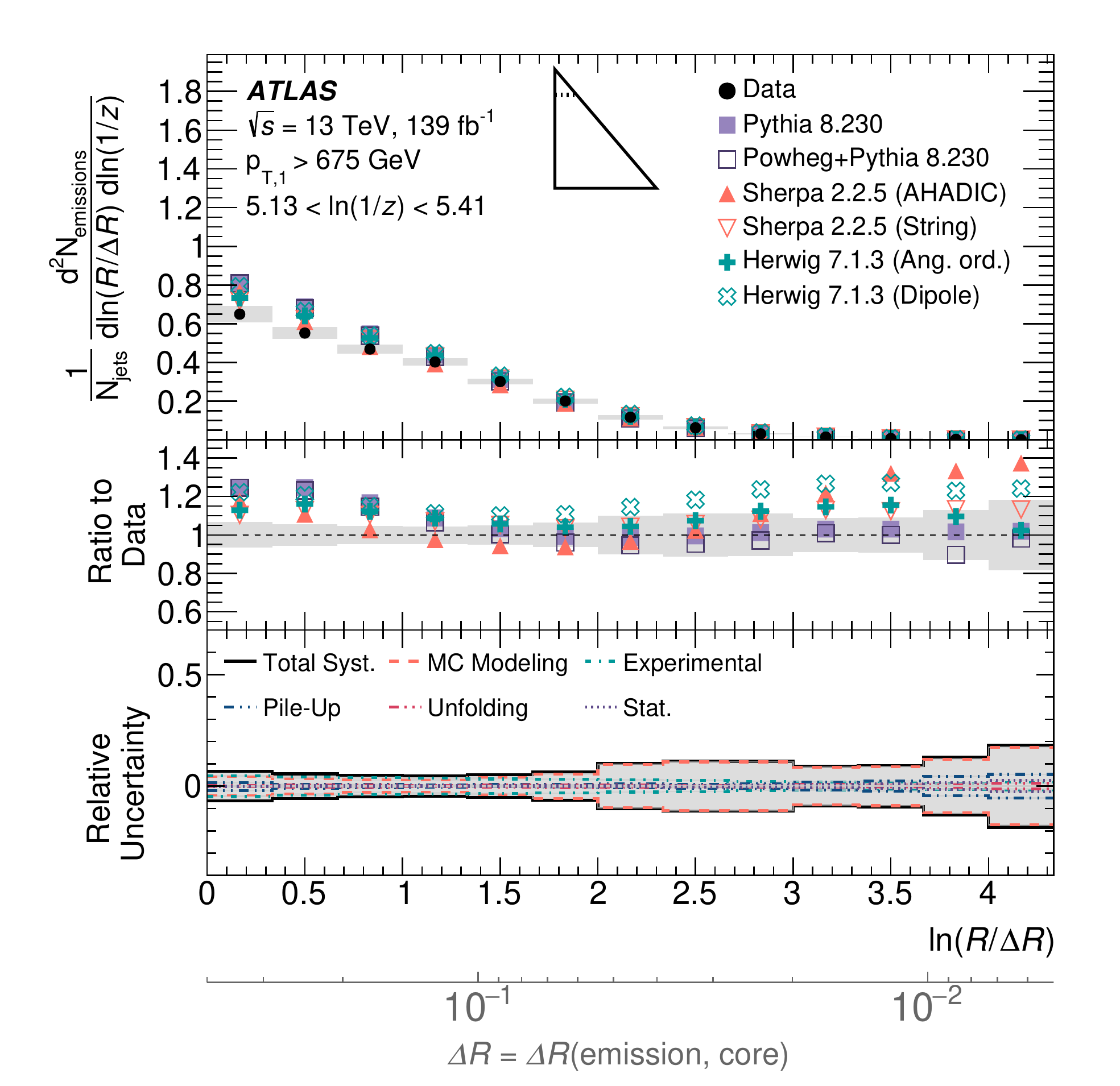}\label{fig:results:slices:d}}
\caption{
Representative horizontal and vertical slices through the LJP. Unfolded data are compared with particle-level simulation from several MC generators.  The uncertainty band includes all sources of systematic and statistical uncertainty.  The inset triangle illustrates which slice of the plane is depicted: (a) $0.67 < \ln(R/\Delta R) < 1.00$, (b) $1.80<\ln(1/z) < 2.08$, (c) $3.33 < \ln(R/\Delta R) < 3.67$, and (d) $5.13<\ln(1/z) < 5.41$.
}
\label{fig:results:slices}
\end{figure}

In summary, a measurement of the jet substructure based on the Lund jet plane is reported. The analysis dataset corresponds to an integrated luminosity of 139~\ifb\ of 13~\TeV\ LHC proton--proton collisions recorded by the ATLAS detector. The measurement is performed on an inclusive selection of dijet events, with a leading jet \pt > 675~\GeV. Selected jets are reconstructed from topological clusters using the \akt algorithm with $R=0.4$, and their associated charged-particle tracks are used to construct the observables of interest. The data are presented as an unfolded double-differential cross section, and compared with several Monte Carlo generators with various degrees of modeling accuracy.  This measurement illustrates the ability of the Lund jet plane to isolate various physical effects, and will provide useful input to both perturbative and nonperturbative model development and tuning.
 
\FloatBarrier
 
\section*{Acknowledgments}
% The next lines are included from the .//acknowledgements/Acknowledgements.tex input file

We thank CERN for the very successful operation of the LHC, as well as the
support staff from our institutions without whom ATLAS could not be
operated efficiently.
 
We acknowledge the support of ANPCyT, Argentina; YerPhI, Armenia; ARC, Australia; BMWFW and FWF, Austria; ANAS, Azerbaijan; SSTC, Belarus; CNPq and FAPESP, Brazil; NSERC, NRC and CFI, Canada; CERN; CONICYT, Chile; CAS, MOST and NSFC, China; COLCIENCIAS, Colombia; MSMT CR, MPO CR and VSC CR, Czech Republic; DNRF and DNSRC, Denmark; IN2P3-CNRS and CEA-DRF/IRFU, France; SRNSFG, Georgia; BMBF, HGF and MPG, Germany; GSRT, Greece; RGC and Hong Kong SAR, China; ISF and Benoziyo Center, Israel; INFN, Italy; MEXT and JSPS, Japan; CNRST, Morocco; NWO, Netherlands; RCN, Norway; MNiSW and NCN, Poland; FCT, Portugal; MNE/IFA, Romania; MES of Russia and NRC KI, Russia Federation; JINR; MESTD, Serbia; MSSR, Slovakia; ARRS and MIZ\v{S}, Slovenia; DST/NRF, South Africa; MINECO, Spain; SRC and Wallenberg Foundation, Sweden; SERI, SNSF and Cantons of Bern and Geneva, Switzerland; MOST, Taiwan; TAEK, Turkey; STFC, United Kingdom; DOE and NSF, United States of America. In addition, individual groups and members have received support from BCKDF, CANARIE, Compute Canada and CRC, Canada; ERC, ERDF, Horizon 2020, Marie Sk{\l}odowska-Curie Actions and COST, European Union; Investissements d'Avenir Labex, Investissements d'Avenir Idex and ANR, France; DFG and AvH Foundation, Germany; Herakleitos, Thales and Aristeia programmes co-financed by EU-ESF and the Greek NSRF, Greece; BSF-NSF and GIF, Israel; CERCA Programme Generalitat de Catalunya and PROMETEO Programme Generalitat Valenciana, Spain; G\"{o}ran Gustafssons Stiftelse, Sweden; The Royal Society and Leverhulme Trust, United Kingdom.
 
The crucial computing support from all WLCG partners is acknowledged gratefully, in particular from CERN, the ATLAS Tier-1 facilities at TRIUMF (Canada), NDGF (Denmark, Norway, Sweden), CC-IN2P3 (France), KIT/GridKA (Germany), INFN-CNAF (Italy), NL-T1 (Netherlands), PIC (Spain), ASGC (Taiwan), RAL (UK) and BNL (USA), the Tier-2 facilities worldwide and large non-WLCG resource providers. Major contributors of computing resources are listed in Ref.~\cite{ATL-GEN-PUB-2016-002}.
 
% End of text imported from the .//acknowledgements/Acknowledgements.tex input file

\clearpage
 
\clearpage
\appendix
 
\printbibliography

\clearpage % ATLAS Collaboration author list
% Reference date of STDM-2018-57 is 2019-07-08
% Author list last updated on date 07-APR-20
% Data extracted on 07-Apr-2020 for paper reference STDM-2018-57
% at 3:49pm
 
\begin{flushleft}
\hypersetup{urlcolor=black}
{\Large The ATLAS Collaboration}

\bigskip

\AtlasOrcid[0000-0002-6665-4934]{G.~Aad}$^\textrm{\scriptsize 102}$,    
\AtlasOrcid[0000-0002-5888-2734]{B.~Abbott}$^\textrm{\scriptsize 129}$,    
\AtlasOrcid{D.C.~Abbott}$^\textrm{\scriptsize 103}$,    
\AtlasOrcid[0000-0002-2788-3822]{A.~Abed~Abud}$^\textrm{\scriptsize 71a,71b}$,    
\AtlasOrcid{K.~Abeling}$^\textrm{\scriptsize 53}$,    
\AtlasOrcid{D.K.~Abhayasinghe}$^\textrm{\scriptsize 94}$,    
\AtlasOrcid[0000-0002-8496-9294]{S.H.~Abidi}$^\textrm{\scriptsize 167}$,    
\AtlasOrcid[0000-0002-8279-9324]{O.S.~AbouZeid}$^\textrm{\scriptsize 40}$,    
\AtlasOrcid{N.L.~Abraham}$^\textrm{\scriptsize 156}$,    
\AtlasOrcid[0000-0001-5329-6640]{H.~Abramowicz}$^\textrm{\scriptsize 161}$
% ...
\bigskip
\\

% ...
$^{40}$Niels Bohr Institute, University of Copenhagen, Copenhagen; Denmark.\\
% ...
$^{53}$II. Physikalisches Institut, Georg-August-Universit\"{a}t G\"ottingen, G\"ottingen; Germany.\\
% ...
$^{71}$$^{(a)}$INFN Sezione di Pavia;$^{(b)}$Dipartimento di Fisica, Universit\`a di Pavia, Pavia; Italy.\\
% ...
$^{94}$Department of Physics, Royal Holloway University of London, Egham; United Kingdom.\\
% ...
$^{102}$CPPM, Aix-Marseille Universit\'e, CNRS/IN2P3, Marseille; France.\\
$^{103}$Department of Physics, University of Massachusetts, Amherst MA; United States of America.\\
% ...
$^{129}$Homer L. Dodge Department of Physics and Astronomy, University of Oklahoma, Norman OK; United States of America.\\
% ...
$^{156}$Department of Physics and Astronomy, University of Sussex, Brighton; United Kingdom.\\
% ...
$^{161}$Raymond and Beverly Sackler School of Physics and Astronomy, Tel Aviv University, Tel Aviv; Israel.\\
% ...
$^{167}$Department of Physics, University of Toronto, Toronto ON; Canada.\\
% ...
$^{a}$ Also at Borough of Manhattan Community College, City University of New York, New York NY; United States of America.\\
$^{b}$ Also at CERN, Geneva; Switzerland.\\
% ...

\end{flushleft}

% Created with Glance <Atlas.Glance@cern.ch>

\end{document}